\documentclass[preprint]{aastex}

\slugcomment{Astrophysical Journal Letter}

\shorttitle{UGC 5101}
\shortauthors{Imanishi et al.}

\begin{document}

\title{Strong Evidence for a Buried AGN in UGC 5101: 
Implications for LINER-Type Ultra-Luminous Infrared Galaxies}

\author{Masatoshi Imanishi\altaffilmark{1}}
\affil{National Astronomical Observatory, Mitaka, Tokyo 181-8588, Japan}
\email{imanishi@optik.mtk.nao.ac.jp} 

\author{C. C. Dudley\altaffilmark{2}}
\affil{Naval Research Laboratory, Remote Sensing Division, 
Code 7217, Building 2, Room 240B, 4555 Overlook Ave SW, 
Washington DC 20375-5351, U.S.A.}
\email{dudley@vivaldi.nrl.navy.mil}

\and 

\author{Philip R. Maloney}
\affil{Center for Astrophysics and Space Astronomy, 
University of Colorado, Boulder, CO 80309-0839, U.S.A.}
\email{maloney@casasrv.colorado.edu}
       
\altaffiltext{1}{ Visiting Astronomer at the Infrared Telescope Facility,
which is operated by the University of Hawaii under contract from the 
National Aeronautics and Space Administration.}

\altaffiltext{2}{NRL/NRC Research Associate}

\begin{abstract}

We report on the results of 3--4 $\mu$m spectroscopy of the
ultra-luminous infrared galaxy (ULIRG) UGC 5101.  It has a cool
far-infrared color and a LINER-type optical spectrum, and so, based on a
view gaining some currency, would be regarded as dominated by star
formation.  However, we find that it has strong 3.4 $\mu$m
carbonaceous dust absorption, low-equivalent-width 3.3 $\mu$m
polycyclic aromatic hydrocarbon (PAH) emission, and a small 3.3 $\mu$m
PAH to far-infrared luminosity ratio. This favors an alternative
scenario, in which an energetically dominant AGN is present behind
obscuring dust.  The AGN is plausibly obscured along all lines of
sight (a `buried AGN'), rather than merely obscured along our
particular line of sight.  Such buried AGNs have previously been found
in thermal infrared studies of the ULIRGs IRAS 08572+3915 and 
IRAS F00183$-$7111, both classified optically as LINERs.  
We argue that buried AGNs can produce LINER-type optical spectra, and that
at least some fraction of LINER-type ULIRGs are predominantly powered by
buried AGNs.

\end{abstract}


\section{Introduction}

Ultra-luminous infrared galaxies (ULIRGs) radiate most of their
extremely large, quasar-like luminosities ($> 10^{12}L_{\odot}$) as
infrared dust emission, and dominate the bright end of the galaxy
luminosity function in the nearby universe \citep{soi87}.  Recent
studies have revealed that the bulk of the cosmic sub-mm background
emission has been resolved into discrete sources, similar to nearby
ULIRGs \citep{bla99a}, and for this reason data on nearby ULIRGs have
been extensively used to derive information on star-formation rates,
dust content, and metallicity in the early universe
\citep{bar00,bla99b}.  Understanding the nature of nearby ULIRGs is
therefore of particular importance both locally and cosmologically.
                                                                       
In contrast to the majority of less infrared luminous 
($<$ 10$^{12}L_{\odot}$) galaxies, it has been suggested that 
energetically important, dust-obscured AGNs are present in ULIRGs 
(Veilleux, Kim, \& Sanders 1999a; Fischer 2000).
If the geometry of the dust distribution is toroidal and/or the amount of
dust along our line of
sight is not too great, signatures of such dust-obscured AGNs can be
found in the optical and/or near-infrared wavelength range at 
$\lambda <$ 2 $\mu$m \citep{vei99a,vei99b}.  
However, detecting AGNs that are deeply embedded in a spherical dust shell
(hereafter buried AGNs) and estimating 
their energetic importance is very difficult at $\lambda <$ 2 $\mu$m, 
even though they are energetically important, since spectral
tracers are dominated by less strongly obscured star-formation activity.  
To detect such buried AGNs and estimate their energetic
importance, we must observe at wavelengths where the effects of dust
extinction are smaller.

At wavelengths 3--13 $\mu$m, dust extinction is lower. 
Furthermore, emission is dominated by radiation from dust, the dominant
emission mechanism of ULIRGs.
Thus, discussions of the energy sources of ULIRGs are more straightforward
than those based on observations at X-ray or radio wavelengths, which are
also potentially capable of detecting buried AGNs, but are dominated by
emission mechanisms other than dust emission.
Finally, by using spectral features at 3--13 $\mu$m, we can distinguish the
energy sources of individual galaxies: while star-formation-dominated
galaxies show strong polycyclic aromatic hydrocarbon (PAH) emission
features, buried AGNs should show dust absorption features below
smooth continuum emission \citep{roc91,gen98,dud99,imd00}.  Using
spectral features, particularly the 7.7 $\mu$m PAH emission, in the
rest-frame 5--11 $\mu$m spectra obtained with the {\it Infrared Space
Observatory} (ISO), it has been argued that the majority of ULIRGs are
star-formation powered \citep{gen98,rig99,tra01}.  However, the
determination of the continuum level with respect to which emission
and absorption features should be measured in {\it ISO} spectra is
highly uncertain, due to insufficient wavelength coverage longward of
these features. 
Distinguishing between star-formation and buried AGN activity therefore
remains controversial \citep{dud99,spo01}. 

As discussed in \citet{imd00}, observations at 3--4 $\mu$m have two
important advantages: firstly, dust extinction is as small as that at
7--8 $\mu$m \citep{lut96}; and secondly, the 3.3 $\mu$m PAH emission
and 3.4 $\mu$m carbonaceous dust absorption features can be used to
distinguish between the different energy sources of ULIRGs, without
serious uncertainty in the continuum determination.  Ground-based 3--4
$\mu$m spectroscopy is thus potentially a powerful tool to investigate
the energetic importance of AGNs buried in the compact nuclei of
ULIRGs, and to settle the controversy discussed above.  Our particular
interest is the search for buried AGNs in ULIRGs with cool
far-infrared colors and/or non-Seyfert optical spectra, which are
typically taken to be star-formation-dominated, based on {\it ISO}
studies (Genzel \& Cesarsky 2000; Taniguchi et al. 1999; Lutz,
Veilleux, \& Genzel 1999).  UGC 5101 is such a ULIRG (Table 1 in this  
paper; Veilleux et al. 1999a), and has actually been diagnosed as being 
star-formation powered \citep{rig99}.  
In this Letter, we report on the results of 3--4 $\mu$m spectroscopic  
observations of UGC 5101, and the discovery of evidence for an  
energetically important buried AGN in this source.  
Throughout this paper, $H_{0}$ $=$ 75 km s$^{-1}$ Mpc$^{-1}$,
$\Omega_{\rm M}$ = 0.3, and $\Omega_{\rm \Lambda}$ = 0.7 are adopted.

\section{Observations and Data Analysis}

We used the NSFCAM grism mode \citep{shu94} to obtain a 3--4 $\mu$m 
spectrum of UGC 5101 at IRTF on Mauna Kea, Hawaii on the night of 2001
April 9 (UT).
Sky conditions were photometric throughout the observations, and 
the seeing was measured to be 0$\farcs$7--0$\farcs$8 in full-width at 
half maximum.
The detector was a 256 $\times$ 256 InSb array.
The HKL grism and L blocker were used with the 4-pixel slit (=
1$\farcs2$).
The resulting spectral resolution was $\sim$150 at 3.5 $\mu$m.

The spectrum of UGC 5101 was obtained toward the flux peak at 
3--4 $\mu$m.
The position angle of the slit was 0 degree east of north.
A standard telescope nodding technique with a throw of 12$''$   
was employed along the slit to subtract background emission.
HR 4112 (F8V, V=4.8) was observed with an airmass difference of 
$<$ 0.1 to correct for the transmission of the Earth's atmosphere, and
provide flux calibration.

Standard data analysis procedures were employed, using IRAF   
\footnote{
IRAF is distributed by the National Optical Astronomy Observatories, 
which are operated by the Association of Universities for Research 
in Astronomy, Inc. (AURA), under cooperative agreement with the 
National Science Foundation.}. 
Initially, bad pixels were replaced with the interpolated signals of
the surrounding pixels.  Bias was subsequently subtracted from the
obtained frames and the frames were divided by a spectroscopic flat
frame.  Finally the spectra of UGC 5101 and HR 4112 were extracted.
Wavelength calibration was performed taking into account the
wavelength-dependent transmission of the Earth's atmosphere.  Since we
set each exposure time to 1.2 sec for the UGC 5101 observation, to
reduce observing overheads, data at $>$ 4 $\mu$m were affected by the
non-linear response of the detector. We excluded these data from our
analysis.  The UGC 5101 spectrum was then divided by the observed
spectrum of HR 4112 and multiplied by the spectrum of a blackbody with
a temperature appropriate to F8V stars (6200K).  By adopting $L$ = 3.4
for HR 4112 based on $V-L$ = 1.4 \citep{tok00}, a flux-calibrated
spectrum was produced.

\section{Results}

\begin{figure}
\plotone{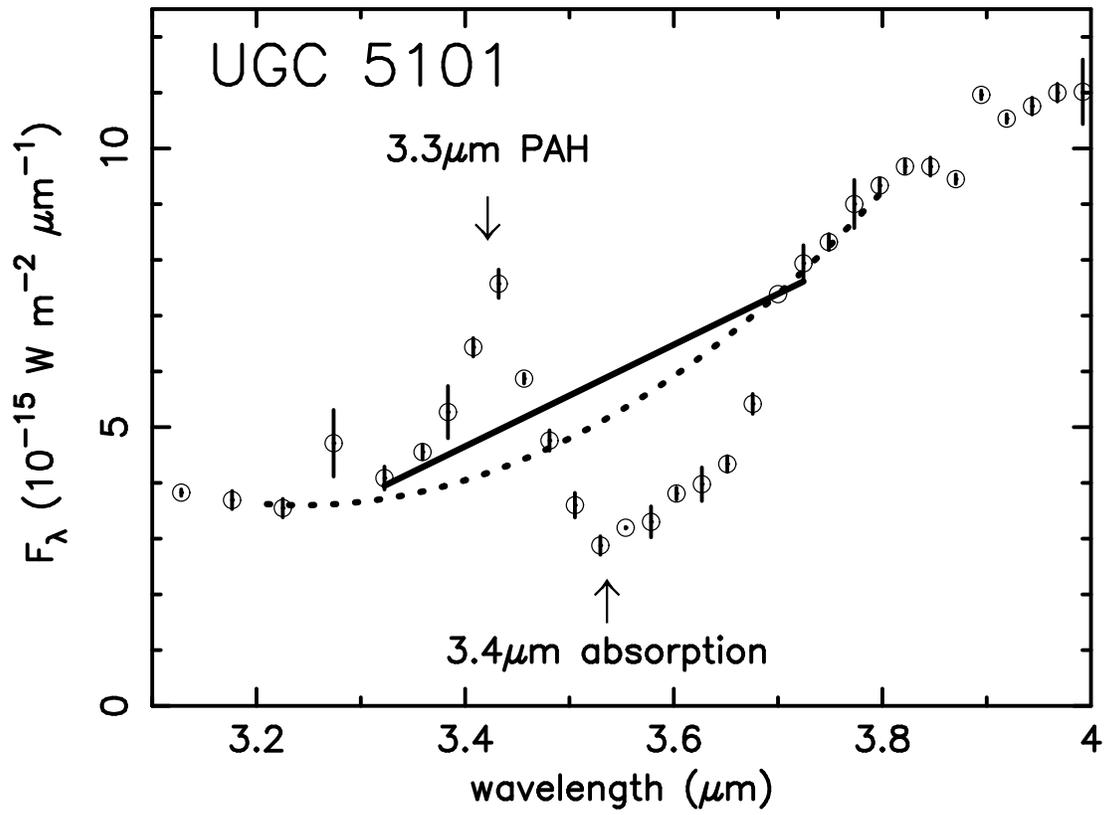}
\caption{3--4 $\mu$m spectrum of UGC 5101. 
The abscissa and ordinate are the observed wavelength and
the flux in F$_{\lambda}$, respectively.
The solid and dashed lines are possible continuum levels (see $\S$ 3). 
\label{fig1}}
\end{figure}

A flux-calibrated 3--4 $\mu$m spectrum of UGC 5101 is shown in
Figure~\ref{fig1}.  At 3--4 $\mu$m, UGC 5101 was spatially very
compact, and its spatial extent along the slit direction was
indistinguishable from a point source.
The spatially-unresolved red near-infrared nucleus reported by 
\citet{sco00} and interpreted as a possible AGN by these authors is the
probable origin of the bulk of the detected 3--4 $\mu$m continuum in our
spectrum. 
Our spectrum gives a value of $L$ (3.55 $\mu$m) of 10.1 mag, which is
similar to the value of $L'$ (3.7 $\mu$m) of 9.8 mag measured with a 5$''$
aperture \citep{san88a}.
Given that UGC 5101 has a red near-infrared color ($K - L'$ = 1.3;
Sanders et al. 1988a), so that $L - L'$ should be $>$ 0 mag, our slit
loss for the continuum emission is less than 0.3 mag.
The dust continuum emission at 10--20 $\mu$m \citep{soi00}, 
Pa$\alpha$ emission (Genzel et al. 1998, the top-middle of Fig 8), 
and 6--18 cm radio continuum emission \citep{sop91,cra96} all show a 
centrally peaked morphology, with spatially-extended emission presumably
originating in weakly obscured star-formation activity. 
Our 1$\farcs$2 slit covers the bulk of this emission, so that most of the 
PAH emission should also be covered with our slit.
Any missing PAH flux is very likely to be much smaller than the detected
PAH flux, and will not seriously affect our quantitative discussion of the
energy source of UGC 5101 ($\S$ 4.1).

The spectrum in Fig.~\ref{fig1} shows a remarkable deviation at
3.3--3.7 $\mu$m from the smooth continuum emission. 
(A linear continuum level is shown as the solid line in Fig.~\ref{fig1}.)
We attribute the emission feature to 3.3
$\mu$m PAH emission, since the flux peak at 3.42 $\mu$m is
consistent with the expected peak wavelength of the redshifted PAH 
emission (3.29 $\mu$m $\times$ 1.040).
Based on our adopted linear continuum level, the observed flux and rest-frame 
equivalent width of the 3.3 $\mu$m PAH emission are estimated to be 
f$_{\rm PAH}$ = 1.3 $\times$ 10$^{-13}$ ergs s$^{-1}$ cm$^{-2}$ and 
EW$_{\rm PAH}$ = 0.025 $\mu$m, respectively.

At wavelengths longward of the PAH emission, signals at 3.5--3.7
$\mu$m are suppressed compared to the continuum level.  The flux level
suddenly and steeply decreases at $\sim$3.7 $\mu$m, compared to the 
extrapolation from longer wavelengths, so that a strong absorption feature 
is undoubtedly present.  We attribute this flux suppression to the 3.4
$\mu$m carbonaceous dust absorption \citep{pen94}, because the observed
maximum of absorption at $\sim$3.53 $\mu$m is consistent with the
redshifted absorption peak wavelength of the carbonaceous dust
absorption feature (3.4 $\mu$m $\times$ 1.040).  Its observed optical
depth is $\tau_{3.4}$(observed) $\sim$ 0.7.

The 3.3 $\mu$m PAH emission profile almost fades out at 3.33--3.34
$\mu$m \citep{tok91,imd00}, while the 3.4 $\mu$m carbonaceous dust
absorption profile is just beginning at this wavelength range
\citep{pen94}.  At {\it z} = 0.040, this wavelength range, where
neither emission nor absorption is particularly important, is
redshifted to 3.46--3.47 $\mu$m.  It will be seen that in
Fig.~\ref{fig1} the adopted linear continuum intersects the observed
data points at this wavelength, providing further evidence that the
continuum determination is reasonable.  An alternative hypothesis is
represented by the curved continuum shown as the dashed line in
Fig.~\ref{fig1}. If this model were adopted, the observed flux and rest frame
equivalent width of the 3.3 $\mu$m PAH emission would increase to
f$_{\rm PAH}$ = 2.0 $\times$ 10$^{-13}$ ergs s$^{-1}$ cm$^{-2}$ and
EW$_{\rm PAH}$ = 0.045 $\mu$m, respectively, and the observed optical
depth of the 3.4 $\mu$m dust absorption would decrease to
$\tau_{3.4}$(observed) $\sim$ 0.6.  However, the level of this curved
continuum seems to be too low; it requires that the absorption optical
depth be close to zero at 3.48 $\mu$m or 3.35 $\mu$m in the rest
frame, where the optical depth should in fact be significant
\citep{pen94}.  We thus believe that the actual PAH flux and
equivalent width are lower than the values inferred using the curved
continuum level; we can use these latter values as stringent upper
limits.  Considering the uncertainty of the continuum determination,
we combine the values measured based on the two continuum levels and
adopt f$_{\rm PAH}$ = 1.6$\pm$0.3 $\times$ 10$^{-13}$ ergs s$^{-1}$
cm$^{-2}$, EW$_{\rm PAH}$ = 0.035$\pm$0.010 $\mu$m, and
$\tau_{3.4}$(observed) = 0.65$\pm$0.05.

\section{Discussion}

\subsection{A buried AGN in UGC 5101}

The detection of 3.3 $\mu$m PAH emission indicates the presence of 
star-formation activity.
However, its observed luminosity is 
5.2$\pm$1.0 $\times$ 10$^{41}$ ergs s$^{-1}$,
which yields an observed 3.3 $\mu$m PAH to far-infrared luminosity 
ratio (Table 1) of $\sim$1 $\times$ 10$^{-4}$, an order of magnitude
smaller than is found in star-formation-dominated galaxies 
($\sim$1 $\times$ 10$^{-3}$; Mouri et al. 1990).
The detected, weakly-obscured star-formation activity thus contributes little
to the huge far-infrared luminosity of UGC 5101 and a dominant energy
source must be located behind the dust.

The presence of such a dust-obscured energy source is supported by the 
detection of strong 3.4 $\mu$m carbonaceous dust absorption. 
If the 3--4 $\mu$m continuum emission source behind the dust originated in 
star-formation activity, the rest-frame equivalent width of the 
3.3 $\mu$m PAH 
emission should be similar to those of less obscured 
star-formation-dominated galaxies 
($\sim$0.12 $\mu$m; Imanishi \& Dudley 2000), 
because both 3--4 $\mu$m continuum and 3.3 $\mu$m PAH emission fluxes
would be attenuated similarly by dust extinction.
The observed rest-frame equivalent width is, however, smaller by more than
a factor of three.
Consequently, the emission from behind the dust shows virtually
no PAH emission, suggesting that the source is an AGN.
This AGN activity dominates the observed 3--4 $\mu$m
continuum flux, contributing $\sim$70 \% of it, and is plausibly the
energy source of UGC 5101's far-infrared luminosity.
If the dust obscuring such an energetically important AGN had a 
torus-like geometry, we would expect UGC 5101 to be optically classified as 
a Seyfert 2, 
but in fact the non-Seyfert optical classification of UGC 5101 \citep{vei99a} 
implies that the AGN is obscured by dust along all lines of
sight (that is, a buried AGN). 
                                   
After subtracting the contribution from star-formation to the observed 
3--4 $\mu$m fluxes by using the spectral shape of the starburst
galaxy NGC 253 in \citet{imd00}, we estimate the intrinsic optical
depth of the 3.4 $\mu$m carbonaceous dust absorption toward the buried
AGN to be $\tau_{3.4}$(intrinsic) $\sim$ 0.8, which yields dust
extinction of $A_{\rm V}$ $>$ 100 mag if a Galactic extinction curve
is assumed \citep{pen94}.  Assuming A$_{3-4 \mu m}$ $\sim$ 0.05
$\times$ A$_{\rm V}$ \citep{lut96}, we estimate the dereddened 3--4
$\mu$m dust emission luminosity  powered by the buried
AGN to be $\nu$L$_{\nu}$ 
${^{\displaystyle >}_{\displaystyle \sim}}$10$^{45}$ ergs s$^{-1}$, 
which is comparable to the
{\it observed} peak dust emission luminosity at 60 $\mu$m and 100
$\mu$m ($\nu$L$_{\nu}$ $\sim$ 2 $\times$ 10$^{45}$ ergs s$^{-1}$, as
calculated based on the {\it IRAS} fluxes shown in Table 1).  In the
case of a buried AGN, dust radiative transfer controls the
temperature of the dust shell; the temperature of the dust
decreases with increasing distance from the central AGN.  The entire
luminosity is transferred at each temperature.  The present data thus
provide evidence for an inner $\sim$900 K, 3--4 $\mu$m continuum
emitting dust shell of a luminosity similar to that of the observed
outer $\sim$40 K, 60--100 $\mu$m continuum emitting dust shell, as is
expected from the buried AGN scenario.
             
For obscured AGNs, \citet{alo97} found that the column density of X-ray
absorbing gas, parameterized by $N_{\rm H}$, relative to dust extinction
towards the 3--4 $\mu$m continuum emitting region ($A_{\rm V}$) is often
higher by a large factor than 
the Galactic $N_{\rm H}$/$A_{\rm V}$ ratio (1.8 $\times$ 10$^{21}$
cm$^{-1}$ mag$^{-1}$; Predehl \& Schmitt 1995).  Dust obscuration
toward the 3--4 $\mu$m continuum emitting region around the buried AGN  
in UGC 5101 is so high ($A_{\rm V} >$ 100 mag: this work) that
$N_{\rm H}$ could easily exceed 10$^{24}$ cm$^{-2}$, in which case
direct 2--10 keV X-ray emission from the buried AGNs would be
very strongly attenuated. 
The non-detection of 2--10 keV X-rays from the putative buried AGN 
with ASCA \citep{nak99} can thus be explained without difficulty.

The radio to far-infrared flux ratio of UGC 5101 is a factor of 
two larger than the typical values for star-formation-dominated
galaxies at 1.5 GHz \citep{cra96}, but inside their scatter 
at 151 MHz and 5 GHz \citep{cox88,sop91}. 
Though the bulk of the radio emission is extended \citep{cra96,sop91}, 
the spatially-unresolved VLBI radio core, interpreted as an AGN by 
\citet{smi98}, is luminous enough to account for the bolomotric
luminosity of UGC 5101 with AGN activity \citep{lon95}. 

\subsection{Implications for optically LINER-type ULIRGs}

Besides UGC 5101, studies in the thermal infrared have provided 
evidence for the presence of energetically dominant buried AGNs 
in two other optically non-Seyfert ULIRGs, 
IRAS 08572+3915 \citep{daw97, imd00} and IRAS F00183$-$7111 \citep{tra01}. 
The far-infrared emission properties of these three ULIRGs are summarized 
in Table 1. 
In the optical, all are classified as LINERs \citep{vei99a,lut99}.
For ULIRGs classified optically as LINERs, which constitute $\sim$40\% of
ULIRGs \citep{vei99a}, two arguments based on {\it ISO} studies have been
made: 
(1) star-formation activity is energetically dominant, and 
(2) the LINER-type optical line emission is due to superwind-driven shocks
caused by star-formation activity \citep{tan99,lut99}. 
The first argument, however, is not applicable to the three ULIRGs
in Table 1. 
Then what is the origin of the LINER-type optical emission
in these objects?
                                        
In buried AGNs, UV emission is blocked at the inner surface of the
surrounding dust shell, but X-rays can penetrate deeply into the dust
to produce X-ray dissociation regions (XDRs; Maloney, Holllenbach, \&
Tielens 1996).  In XDRs, locally-generated UV photons, such as
Lyman-Werner band H$_{2}$ emission and Ly$\alpha$ emission, can ionize
some species with ionization potential less than 10--11 eV (notably
carbon and iron), but the emitting volume is dominated by largely
neutral gas, which causes XDRs essentially always to have LINER-type
spectra (Maloney, in preparation).  
A buried AGN can thus explain the observed LINER-type optical emission.

However, in practical cases, some star-formation activity may very
well be present at the surface of galaxies even when buried AGNs are
energetically dominant, as we infer for UGC 5101.  
In such a case, due to the susceptibility of optical emission
to dust extinction, the optical spectral diagnostics are likely to be 
dominated by this surface star-formation emission.  
The star-formation activity could be 
responsible for LINER-type optical emission via shocks, or might even
alter the optical emission to a HII region type.  
It is important to note that narrow-line emission from high-excitation
clouds characteristic of Seyfert galaxy spectra which are observed to be
produced along the axes of nuclear dust tori would not contribute
substantially at any wavelength in the case of buried AGN. 
                                   
\section{Summary}

We have found evidence for the presence of an energetically
dominant, buried AGN in the ULIRG UGC 5101, which is characterized by 
a cool far-infrared color and a LINER optical spectrum.  
This finding seems contrary to the
currently established view, based on {\it ISO} data, that such
objects are powered by star formation.  Two other LINER ULIRGs
also show evidence for energetically dominant buried AGNs, and we
therefore conclude that at least some fraction of LINER ULIRGs
are powered by buried AGNs.

\acknowledgments

We are grateful to J. Rayner and B. Golisch for their support before
and during the observing run.
Research in infrared astronomy at NRL is supported by the Office of Naval
Research.
PRM is supported by the NSF under grant AST-9900871.
This research has made use of the NASA/IPAC Extragalactic Database
(NED) which is operated by the Jet Propulsion Laboratory, California
Institute of Technology, under contract with the National Aeronautics
and Space Administration.

\clearpage

\begin{deluxetable}{lcrrrcl}
\tablecaption{Summary of UGC 5101, IRAS 08572+3915, and IRAS 
F00183$-$7111. \label{tab1}}
\tablewidth{0pt}
\tablehead{
\colhead{Object} & \colhead{Redshift}   & 
\colhead{f$_{\rm 25}$ \tablenotemark{a}}   & 
\colhead{f$_{\rm 60}$ \tablenotemark{a}}   & 
\colhead{f$_{\rm 100}$ \tablenotemark{a}}  & 
\colhead{log L$_{\rm FIR}$ \tablenotemark{b}} & 
\colhead{f$_{25}$/f$_{60}$ \tablenotemark{c}}  \\
\colhead{} & \colhead{}   & \colhead{[Jy]} & \colhead{[Jy]} 
& \colhead{[Jy]}  & \colhead{[ergs s$^{-1}$]} & \colhead{} 
}
\startdata
UGC 5101 & 0.040 & 1.03 & 11.54 & 20.23 & 45.61 & 0.09 (cool)\\
IRAS 08572+3915 & 0.058 & 1.70 & 7.43 & 4.59 & 45.62 & 0.23 (warm)\\
IRAS F00183$-$7111 & 0.327 & 0.13 & 1.20 & 1.19 & 46.51 & 
0.11 (cool)\\ 
\enddata

\tablenotetext{a}{f$_{25}$, f$_{60}$, and f$_{100}$ are 
{\it IRAS FSC} fluxes at 25$\mu$m, 60$\mu$m, and 100$\mu$m, respectively.} 

\tablenotetext{b}{Logarithm of far-infrared (40--500 $\mu$m) luminosity 
            in ergs s$^{-1}$ calculated with
            $L_{\rm FIR} = 1.4 \times 2.1 \times 10^{39} \times$ 
            $D$(Mpc)$^{2}$ $\times (2.58 \times f_{60} + f_{100}$) 
            ergs s$^{-1}$ \citep{sam96}.}

\tablenotetext{c}{f$_{25}$/f$_{60}$ $<$ ($>$) 0.2 are called cool
(warm) \citep{san88b}.}

\end{deluxetable}

\clearpage

\end{document}